\documentclass[aps,twocolumn,showpacs]{revtex4}
\usepackage{graphicx}
\include{amssym}

\begin{document}
\title{The anomalous decay $f_1(1285)\to\rho\gamma$ and related processes}
\author{A. A. Osipov\footnote{Email address: osipov@nu.jinr.ru},
             A. A. Pivovarov\footnote{Email address: tex\_k@mail.ru},
             M. K. Volkov\footnote{Email address: volkov@theor.jinr.ru}}
\affiliation{Joint Institute for Nuclear Research, Bogoliubov Laboratory of Theoretical Physics, 141980 Dubna, Russia}

\begin{abstract}
We work out the low-energy expansion of the anomalous $f_1(1285)\to\rho\gamma$ decay amplitude by using the Nambu--Jona-Lasinio model with $U(2)\times U(2)$ chiral symmetric four-quark interactions in the one-quark-loop approximation. The related processes $f_1(1285)\to\omega\gamma$, $a_1(1260)\to\omega\gamma$, and $a_1(1260)\to\rho\gamma$, are also considered. An effective meson Lagrangian responsible for $f_1\rho\gamma$, $f_1\omega\gamma$, $a_1\rho\gamma$ and $a_1\omega\gamma$ interactions is found. The predicted radiative decay widths, $\Gamma_{f_1\to\rho^0\gamma}=311\ \mbox{keV}$,  $\Gamma_{f_1\to\omega\gamma}=34.3\ \mbox{keV}$, $\Gamma_{a_1\to\rho^0\gamma}=26.8\ \mbox{keV}$, $\Gamma_{a_1\to\omega\gamma}=238\ \mbox{keV}$, allow an experimental test of the hypothesis that $f_1(1285)$ and $a_1(1260)$-mesons have a quark-antiquark nature. At present, only the $f_1(1285)\to\rho\gamma$ decay has been measured. Our result is in remarkably good agreement with the recent data of CLAS Collaboration $\Gamma_{f_1\to\rho^0\gamma}=453\pm 177\ \mbox{keV}$, but disagrees with the PDG-based estimate of $\Gamma_{f_1\to\rho^0\gamma}=1326\pm 313\ \mbox{keV}$. The calculations presented require a minimum of theoretical input, and are shown to be consistent with the non-renormalization theorems of QCD.  
\end{abstract}

\maketitle

\section{Introduction}
Anomalies have important consequences for a wide range of issues in quantum field theory. This is why they are invariably under special attention of theoreticians and experimentalists. For instance, the Wess-Zumino effective Lagrangian \cite{Wess71} summarizes the effects of anomalies in current algebra and finally relates uniquely the $\pi^0\to\gamma\gamma$ decay amplitude with other ones, such as $\gamma\to 3\pi$, $\gamma\gamma\to 3\pi$ and a five pseudoscalar vertex. Thus, the anomaly based results are tightly restrictive and potentially very accurate. The latter is the consequence of the Adler-Bardeen theorem \cite{Adler69} which states that chiral anomaly is not modified by higher order corrections. 

The world average phenomenological data on the radiative decay $f_1\to\rho^0\gamma$ \cite{PDG} selected through the study of the reaction $\pi^-N\to\pi^-f_1(1285) N\to \pi^-\pi^+\pi^-\gamma N$, and new results of CLAS Collaboration at Jefferson Lab. on the $f_1(1285)$ photoproduction off a proton target \cite{Dickson16} are especially important due to a presence of the anomaly: the underlying triangle quark loop diagrams describing the $f_1\rho^0\gamma$ and $f_1\omega\gamma$ vertices. These vertices determine the widths of the $f_1\to\rho^0\gamma$ and $f_1\to\omega\gamma$ decays and are basic elements in the description of the $f_1(1285)$ photoproduction data. In this respect the presently available phenomenological data allow a sensitive test of the $f_1\rho\gamma$ anomaly. Yet some of the essential properties of the theoretical description of this vertex are only poorly understood.

The connections of vector and axial-vector mesons with the anomaly can be studied on general grounds, i.e without assuming the quark-antiquark structure for spin-1 states. For instance, the method, based on the massive Yang-Mills approach \cite{Schechter84}, leads to the Bardeen's form of the non-abelian anomaly. Unfortunately, this breaks explicitly the chiral $SU(3)_L\times SU(3)_R$ symmetry and forbids the $f_1\rho^0\gamma$, $f_1\omega\gamma$, $a_1\rho\gamma$, and $a_1\omega\gamma$ vertices. 

On the opposite, if one starts from the most general anomalous action in terms of pseudoscalars and spin-1 states which is chirally (gauge) symmetric and embodies the chiral anomalies only through the Wess-Zumino-Witten action of the pseudoscalars \cite{Witten83}, one gets a consistent scheme, and in this case there is the possibility for anomalous $f_1\rho^0\gamma$, $f_1\omega\gamma$, $a_1\rho\gamma$, and $a_1\omega\gamma$ vertices \cite{Kugo85,Meissner90}. However, the method fails to predict the couplings of the effective Lagrangian, and cannot be used to estimate the width of the $f_1\to\rho^0\gamma$ decay.     

The purpose of the paper is to clarify exactly this obscure aspect of the radiative decay $f_1\to\rho^0\gamma$ of the $f_1(1285)$ axial-vector $I^G(J^{PC})=0^+(1^{++}$) meson. For definiteness we will consider the Nambu--Jona-Lasinio (NJL) model with $U(2)\times U(2)$ chiral symmetry spontaneously broken down to the diagonal $SU(2)_I$ subgroup (the quantum anomaly breaks the axial $U(1)$ symmetry) \cite{Ebert83,Volkov84,Volkov86,Ebert86,Ebert94,Osipov94,Bernard96,Volkov06}. The NJL model not only gives the structure of the vertex, but also fixes the values of the coupling constants involved. 

There are at least three essential reasons for our calculations. First, we show that if one assumes that the $f_1(1285)$ meson is a bound quark-antiquark state one can obtain its radiative decay width by considering the anomalous quark triangle diagram. The result is restrictive because there are general statements about the longitudinal and transversal parts of the triangle \cite{Adler69,Vainshtein03}.  

Second, the phenomenological data on this decay presently are very contradictive. The PDG-based estimate is $\Gamma_{f_1\to\rho^0\gamma}=1326\pm 313\ \mbox{keV}$. The recent CLAS data give three times less this value $\Gamma_{f_1\to\rho^0\gamma}=453\pm 177\ \mbox{keV}$. We argue that the quite low value reported by the CLASS collaboration can be perfectly understood if the $f_1$-meson is approximately a quark-antiquark $\bar nn=\frac{1}{\sqrt{2}}(\bar uu +\bar dd)$ state. 

Third, it has been noted in \cite{Wang17} that known theoretical models \cite{Kochelev09,Domokos09} failed to fit new CLAS data. We suppose that one of the reasons is related with the oversimplified expressions used for $f_1\rho^0\gamma$ and $f_1\omega\gamma$ vertices. In this respect a new attempt made in \cite{Wang17} also suffers from a superficial approach to the anomalous structure of these vertices. We suggest a different effective Lagrangian which consistently describes the triangle anomaly and can be applied to fit CLAS data.  

\section{Anomalous triangle diagram for the $f_1\to\rho^0\gamma$ decay}
\label{L}
The relevant vertices of the NJL quark-meson Lagrangian density are
\begin{eqnarray}
\label{lagf}
&&{\cal L}_f =\frac{g_\rho}{2}\bar q\gamma_\alpha\gamma_5qf_1^\alpha, \\
\label{lagrho}
&&{\cal L}_\rho =\frac{g_\rho}{2}\bar q\tau_3\gamma^\beta q\rho_\beta^0, \\
\label{laggamma}
&&{\cal L}_\gamma =e\bar qQ\gamma^\gamma q A_\gamma , \quad Q=\frac{1}{2}\left(\tau_3 +\frac{1}{3}\right), 
\end{eqnarray}
where the $q(x)$ are constituent quarks of mass $m$ (the color and flavor indices are suppressed); $f_1^\alpha(x)$ is an axial-vector $f_1(1285)$ field; $\rho_\beta^0(x)$ is a neutral $\rho (770)$-meson field; $A_\gamma (x)$ is a photon field; $Q$ is the matrix of the light quark's charges, where $\tau_3$ is a diagonal Pauli-matrix; $\gamma^\mu$ are the standard Dirac matrices in four dimensions. The couplings $g_\rho$ and $e$ are well established: $e$ is the proton charge with $\alpha=e^2/(4\pi)=1/137$ and $g_\rho$ is fixed from the $\rho\to\pi\pi$ decay, $\alpha_\rho=g_\rho^2/(4\pi)\simeq 3$. The constituent quark mass $m$ is equal to $m=280\ \mbox{MeV}$ \cite{Volkov86}. 

The amplitude of the $f_1\to\rho^0\gamma$ decay can be written as follows
\begin{equation}
\label{amplitude}
{\cal M}=-iN_c\frac{eg_\rho^2}{16\pi^2}\epsilon_\alpha(l)\epsilon^*_\beta(p)\epsilon_\gamma^*(q)T^{\alpha\beta\gamma}(p,q),
\end{equation}
where $N_c=3$ is a number of colors, $\epsilon_\alpha (l)$ is the polarization vector of the $f_1$-meson, $\epsilon_\beta (p)$ is a polarization vector of $\rho$-meson and $\epsilon_\gamma(q)$ is the polarization vector of the photon; $p$ and $q$ are 4-momenta of $\rho(770)$ and photon correspondingly, $l=p+q$ is the 4-momentum of $f_1$-meson. The tensor $T^{\alpha\beta\gamma}(p,q)$ is a sum of two Feynman diagrams (see fig.1) 

\begin{figure}[htb]
\label{fig1}
\includegraphics[width=0.45\textwidth]{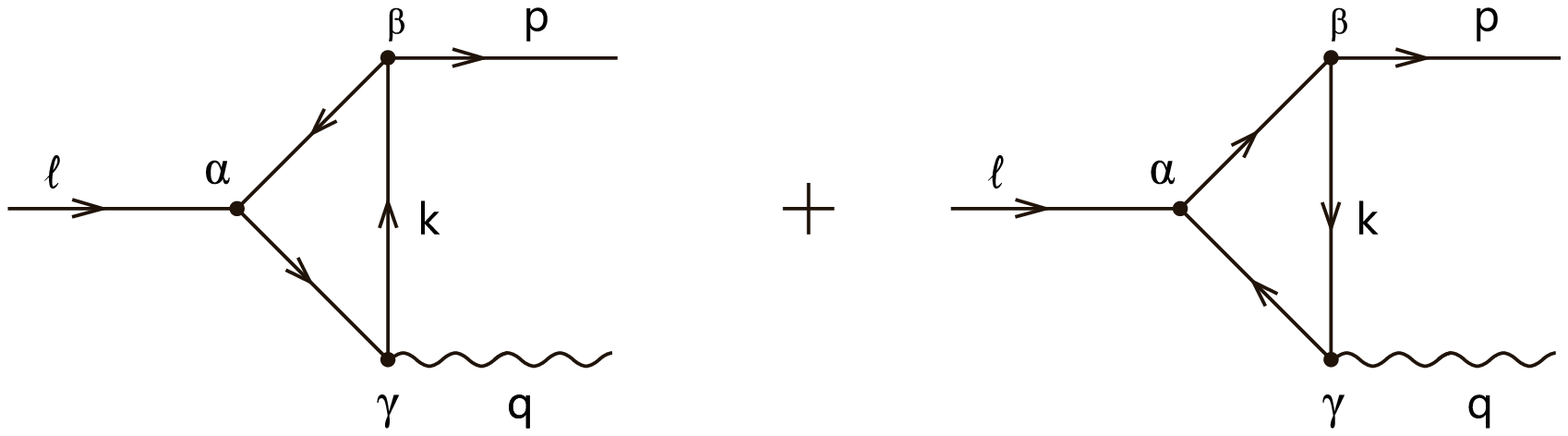}
\caption{The one-quark-loop contribution to the radiative decay amplitude $f_{1\alpha}(l)\to\rho_\beta(p)\gamma_\gamma (q)$. The first diagram corresponds to the Lorentz tensor $\tilde T^{\alpha\beta\gamma}(p,q)$, the second one to the tensor $\tilde T^{\alpha\gamma\beta}(q,p)$. External lines represent the $f_1$ axial-vector field with polarization vector $\epsilon_\alpha (l)$, the vector $\rho$-meson field with polarization vector $\epsilon^*_\beta(p)$, and electromagnetic field $A_\gamma$ with polarization vector $\epsilon^*_\gamma(q)$.}
\end{figure} 

\begin{equation}
\label{ampl}
T^{\alpha\beta\gamma}(p,q)=\tilde T^{\alpha\beta\gamma}(p,q)+\tilde T^{\alpha\gamma\beta}(q,p),
\end{equation}
where 
\begin{eqnarray}
&&\!\!\!\!\!\!\!\!\!\!\!\!\!\!\!\! \tilde T_{\alpha\beta\gamma}(p,q)= \nonumber \\
&&\!\!\!\!\!\!\!\!\!\!\!\!\!\!\!\!\int\!\!\frac{d^4k}{4\pi^2}\frac{\mbox{Tr}[\gamma_5\gamma_\alpha(\hat k-\hat p+m)\gamma_\beta(\hat k+m)\gamma_\gamma(\hat k+\hat q+m)]}{(k^2-m^2)[(k-p)^2-m^2][(k+q)^2-m^2]}. 
\end{eqnarray}
It is obvious from (\ref{ampl}) that the Lorentz tensor $T^{\alpha\beta\gamma}(p,q)$ obeys the Bose symmetry requirement
\begin{equation}
\label{BS}
T^{\alpha\beta\gamma}(p,q)=T^{\alpha\gamma\beta}(q,p).
\end{equation} 
Making a replacement of variables $k_\mu\to -k_\mu$ in one of the integrals in (\ref{ampl}), and calculating traces, we obtain that $\tilde T^{\alpha\beta\gamma}(p,q)=\tilde T^{\alpha\gamma\beta}(q,p)$.
   
After some mildly tedious calculations we find that the amplitude can be written in the form
\begin{eqnarray}
\label{T}
&&T_{\alpha\beta\gamma}(p,q)=e_{\alpha\beta\gamma\sigma} (a+q-p)^\sigma \nonumber \\ 
&& +\frac{1}{6m^2}\left\{ e_{\alpha\beta\gamma\sigma}\left[q^\sigma(qp+2p^2)-p^\sigma(qp+2q^2)\right]\right. \nonumber \\
&&\left. -(q+2p)_\beta e_{\alpha\gamma\rho\sigma}q^\rho p^\sigma +(p+2q)_\gamma e_{\alpha\beta\rho\sigma}q^\rho p^\sigma\right\} \nonumber \\
&&+{\cal O}(p^5),
\end{eqnarray}
where we have restricted ourselves up to the terms of the third power in momenta. (To describe correctly the low-energy limit, the amplitude must have the smallest possible number of momenta. One should not think about this truncation in terms of $p^2/m^2$ expansion which is not applicable here. We are following instead the idea of $1/N$ expansion. According to it, the meson physics in the large $N$ limit is described by the tree diagrams of an effective local Lagrangian, with local vertices and local meson fields \cite{Witten79}. This is exactly what one obtains restricting to the leading in momenta terms of the constituent quark loops. The details of such description of low-energy meson physics in the framework of NJL model are given in \cite{Volkov84,Volkov86,Ebert86}. In particular, our result (\ref{T}) differs from the one obtained by Kaiser and Meissner \cite{Meissner90} only by prescribing the definite values to the corresponding couplings of the effective $AVV$-vertices in accord with the NJL model. In Sect. V we show that the truncated triangle diagrams of Fig.1 do reflect the QCD anomaly structure, in the infrared.) 

The result (\ref{T}) contains an ambiguous surface term, represented by the 4-vector $a_\sigma$. It is a well-known remain of superficial linear divergence of the quark-loop integral \cite{Jackiw72}. Most generally $a_\sigma=aq_\sigma+bp_\sigma$, where $a$ and $b$ are free constants. The property (\ref{BS}) relates these constants $a=-b$. The requirement of gauge symmetry is $q_\gamma T^{\alpha\beta\gamma}(p,q)=0$. That totally fix the constants. Indeed, we get from (\ref{T}) 
\begin{equation}
\label{WI}
q_\gamma T^{\alpha\beta\gamma}(p,q)=q^\gamma e_{\alpha\beta\gamma\sigma}(b-1)p^\sigma=0, 
\end{equation}
and if one takes $b=1$ the Ward identity (\ref{WI}) is obviously fulfilled. We conclude that the surface term is completely fixed by Bose and gauge symmetry requirements, $a_\sigma =p_\sigma-q_\sigma$. As a result, the linear in momenta contribution in (\ref{T}) is zero. 

We can gain some deep understanding of this formula by considering the product $p^\beta T_{\alpha\beta\gamma}(p,q)$. It is easy to find out that the result is zero. It means that if one considers the transition of $\rho^0\to\gamma$ in our amplitude (in accord with the idea of vector meson dominance) the formula (\ref{T}) obeys the additional requirement of gauge symmetry. 

Moreover, the celebrated Landau-Yang theorem \cite{Landau48} states that a massive vector (i.e. spin-1) particle cannot decay into two on-shell massless photons. Let us show that our amplitude does not contradict this general result. Indeed, in a frame where $f_1(1285)$ meson is at rest, we can always choose the direction of the $z$-axis along the spatial part of the photons momenta, i.e. $q=(q_0,0,0,q_0)$ and $p=(q_0,0,0,-q_0)$. The photon polarization vectors $\epsilon^*_\gamma (q)$ and $\epsilon^*_\beta (p)$ are orthogonal to the photon momenta, and thus can be chosen as follows: $\epsilon^*_\gamma (q)=(0,\epsilon^*_1(q),\epsilon^*_2(q),0)$ and $\epsilon^*_\beta (p)=(0,\epsilon^*_1(p),\epsilon^*_2(p),0)$. The polarization vector of a massive $f_1$ meson is given by $\epsilon_\alpha (l)=(0,\epsilon_1(l),\epsilon_2(l),\epsilon_3(l))$. It follows then that
\begin{eqnarray}
&& \epsilon^*_\gamma (q)q^\gamma =0, \quad \epsilon^*_\beta (p)p^\beta =0, \\
&& \epsilon^*_\gamma (q)p^\gamma =0, \quad \epsilon^*_\beta (p)q^\beta =0, \\
&& e_{\alpha\beta\gamma\sigma}\epsilon^*_\gamma (q)\epsilon^*_\beta (p)\epsilon_\alpha (l) (q-p)^\sigma =0.
\end{eqnarray}  
Hence, the amplitude of $f_1\to\gamma\gamma$ decay is equal to zero in the $f_1$ boson rest frame. Since the amplitude is Lorentz invariant, it equals zero in any other frame as well.   

Now, it is not difficult to relate the amplitude (\ref{amplitude}) (with $T^{\alpha\beta\gamma}$ given by (\ref{T})) with the Lagrangian density, describing this anomalous decay and leading to the same amplitude  
\begin{eqnarray}
\label{lag-f}
&&{\cal L}_{f_1\rho^0\gamma}=-\frac{e\alpha_\rho}{8\pi m^2}e^{\mu\nu\alpha\beta} \left(\rho^0_{\mu\nu}F_{\alpha\sigma}\partial^\sigma f_\beta \right. \nonumber \\
&&\left. +\frac{1}{2}f_{\mu\nu}F_\alpha^{\ \,\sigma}\rho^0_{\sigma\beta}+F_{\mu\nu}\partial^\sigma \rho^0_{\sigma\alpha}f_\beta\right).   
\end{eqnarray}
Here the quantities $\rho^0_{\mu\nu}, F_{\mu\nu}, f_{\mu\nu}$ stand for the field strengths associated with neutral vector $\rho (770)$-meson field $\rho^0_\mu$,  $\rho^0_{\mu\nu}=\partial_\mu\rho^0_\nu-\partial_\nu\rho^0_\mu$, electromagnetic field $A_\mu$, $F_{\mu\nu}=\partial_\mu A_\nu-\partial_\nu A_\mu$, and the neutral axial-vector $f_{1}(1285)$ field, $f_{\mu\nu}=\partial_\mu f_\nu -\partial_\nu f_\mu$.  This expression gives a definite meaning to our statement about the oversimplified form of the Lagrangian used in the literature for this vertex. To see the difference it is enough to compare (\ref{lag-f}) with the $f_1\rho^0\gamma$ vertex used, for instance, in \cite{Wang17}, where ${\cal L}_{f_1\rho^0\gamma}=\frac{1}{2}eg_{\rho f_1\gamma}e^{\mu\nu\alpha\beta} F_{\mu\nu}\rho_\alpha^0 f_\beta$.  

On the mass surface of $\rho, \gamma$ and $f_1$ mesons from (\ref{amplitude}) and (\ref{T}) we get 
\begin{eqnarray}
&&{\cal M}=-i\frac{e\alpha_\rho}{8\pi m^2}\epsilon_\alpha(l)\epsilon^*_\beta(p)\epsilon_\gamma^*(q)\left[ e^{\alpha\beta\gamma\sigma}\Delta_\sigma \right. \nonumber \\
&&\left. + p_\rho q_\sigma\left(e^{\alpha\gamma\rho\sigma}q^\beta -e^{\alpha\beta\rho\sigma}p^\gamma\right)\right],
\end{eqnarray}   
where $\Delta_\sigma =(qp+2p^2)q_\sigma -(qp)p_\sigma$. Then it follows that
\begin{equation}
|{\cal M}|^2=\frac{m_\rho^4}{2}\left(\frac{e\alpha_\rho}{4\pi m^2}\right)^2(qp)^2\left(\frac{1}{m_\rho^2}+\frac{1}{m_f^2}\right),
\end{equation}
and the radiative decay width $f_1\to\rho^0\gamma$ is given by
\begin{eqnarray}
\Gamma_{f_1\to\rho^0\gamma}&=&\frac{\alpha\alpha_\rho^2}{6(16\pi)^2m_f^5}\frac{m_\rho^2}{m^4}(m_f^2-m_\rho^2)^3(m_f^2+m_\rho^2)\nonumber \\
&=&311\ \mbox{keV}. 
\end{eqnarray}
This model estimate is in a perfect agreement with the experimental result, given by the CLAS Collaboration: $\Gamma_{f_1\to\rho^0\gamma}=453\pm 177\ \mbox{keV}$, and about four times less the PDG estimate: $\Gamma_{f_1\to\rho^0\gamma}=1326\pm 311\ \mbox{keV}$ \cite{PDG}.  

\section{The related decays}
\label{MEL}

Our purpose now is to describe the related processes, i.e. the radiative decays $f_1(1285)\to\omega\gamma$, $a_1(1260)\to\omega\gamma$, and $a_1(1260)\to\rho\gamma$. Their amplitudes are originated by the same quark-loop integrals as the amplitude (\ref{amplitude}). Therefore, the general factor which comes out from the isotopic trace calculations will be the only difference in the results. Let us remind that for the $f_1\to\rho\gamma$ amplitude this factor is $\mbox{tr}(\tau_3Q)=1$. Now, the corresponding factor in the amplitude $f_1\to\omega\gamma$ is $\mbox{tr}(Q)=1/3$. It gives immediately the Lagrangian density     
\begin{eqnarray}
\label{lag-fomega}
&&{\cal L}_{f_1\omega\gamma}=-\frac{e\alpha_\rho}{24\pi m^2}e^{\mu\nu\alpha\beta} \left(\omega_{\mu\nu}F_{\alpha\sigma}\partial^\sigma f_\beta \right. \nonumber \\
&&\left. +\frac{1}{2}f_{\mu\nu}F_\alpha^{\ \,\sigma}\omega_{\sigma\beta}+F_{\mu\nu}\partial^\sigma \omega_{\sigma\alpha}f_\beta\right),   
\end{eqnarray}
and the radiative $f_1\to\omega\gamma$ decay width 
\begin{eqnarray}
\Gamma_{f_1\to\omega\gamma}&=&\frac{\alpha\alpha_\rho^2m_\omega^2}{6(48\pi)^2m_f^5m^4}(m_f^2-m_\omega^2)^3(m_f^2+m_\omega^2) \nonumber \\
&=&34.3\ \mbox{keV}. 
\end{eqnarray}

The $a_1^0\to\omega\gamma$ amplitude has the factor $\mbox{tr}(\tau_3 Q)=1$. It means that 
\begin{eqnarray}
\label{lag-a10omega}
&&{\cal L}_{a_1^0\omega\gamma}=-\frac{e\alpha_\rho}{8\pi m^2}e^{\mu\nu\alpha\beta} \left(\omega_{\mu\nu}F_{\alpha\sigma}\partial^\sigma a^0_\beta \right. \nonumber \\
&&\left. +\frac{1}{2}a^0_{\mu\nu}F_\alpha^{\ \,\sigma}\omega_{\sigma\beta}+F_{\mu\nu}\partial^\sigma \omega_{\sigma\alpha}a^0_\beta\right),   
\end{eqnarray}
and the radiative $a^0_1\to\omega\gamma$ decay width is equal to 
\begin{eqnarray}
\Gamma_{a_1^0\to\omega\gamma}&=&\frac{\alpha\alpha_\rho^2m_\omega^2}{6(16\pi)^2m_{a_1}^5m^4}(m_{a_1}^2-m_\omega^2)^3(m_{a_1}^2+m_\omega^2) \nonumber \\
&=&238\ \mbox{keV}. 
\end{eqnarray}

The radiative decay $a_1\to\rho\gamma$ has three different channels: $a_1^0\to\rho^0\gamma$, and $a_1^\pm\to\rho^\pm\gamma$. Due to the property $\tilde T^{\alpha\beta\gamma}(p,q)=\tilde T^{\alpha\gamma\beta}(q,p)$, we can sum the traces over Pauli matrices of these two contributions. That gives 
\begin{equation}
\mbox{tr}\left(\vec\tau\vec a_1\{\vec\tau\vec\rho , Q\}\right)=\frac{2}{3}\left(\vec a_1\vec\rho\right).
\end{equation}   
We conclude that each of the three possible modes has a similar amplitude and the same expression for the decay width, i.e. 
\begin{eqnarray}
\label{lag-a1rho}
&&{\cal L}_{a_1\rho\gamma}=-\frac{e\alpha_\rho}{24\pi m^2}e^{\mu\nu\alpha\beta} \left(\vec\rho_{\mu\nu}F_{\alpha\sigma}\partial^\sigma \vec a_\beta \right. \nonumber \\
&&\left. +\frac{1}{2}\vec a_{\mu\nu}F_\alpha^{\ \,\sigma}\vec\rho_{\sigma\beta}+F_{\mu\nu}\partial^\sigma \vec\rho_{\sigma\alpha}\vec a_\beta\right),   
\end{eqnarray}
and $\Gamma_{a_1\to\rho\gamma} =\Gamma_{a_1^0\to\rho^0\gamma}=\Gamma_{a_1^\pm\to\rho^\pm\gamma}$, with
\begin{eqnarray}
\Gamma_{a_1\to\rho\gamma}&=&\frac{\alpha\alpha_\rho^2m_\rho^2}{6(48\pi)^2m_a^5m^4}(m_a^2-m_\rho^2)^3(m_a^2+m_\rho^2) \nonumber \\
&=&26.8\ \mbox{keV}. 
\end{eqnarray}
     
\section{Comparison with other approaches}
\label{cwoa}
No much efforts have been done till now for a theoretical description of the processes considered. Probably, this is related to the very poor experimental information available on these radiative decays. In Table \ref{table} we collect the relatively old estimations made in the framework of the covariant oscillator quark model \cite{Ishida89}. Our results, in general, have a tendency to be twice smaller of that predictions. Let us remind that the isoscalar member of the axial-vector $^3P_1$-nonet ($^{2s+1}L_J$), $f_1(1285)$-meson, has mostly $(\bar uu+\bar dd)/\sqrt{2}$ content, but can mix with the mainly $\bar ss$ isosinglet state (and gluons). The authors of \cite{Ishida89} considered two possible candidates for such partner: the usual candidate, $f_1(1420)$, and the another promising state $f_1(1530)$ 
\begin{equation}
f_1(1285)=\bar nn\cos\phi -(\bar ss)\sin\phi .  
\end{equation}
In the case of the combination of two isoscalar members, $f_1(1285)$ and $f_1(1530)$, the mass formula of their model gives the mixing angle $\phi =21^\circ$. In the case of the other combination, $f_1(1285)$ and $f_1(1420)$, the mass formula gives $\phi\simeq 10^\circ$. One can see that data on $f_1(1285)\to\rho\gamma$ and $f_1(1285)\to\omega\gamma$ modes in that model slightly depend on the mixing angle $\phi$. The tendency is the smaller $\phi$ the more radiative decay width (see Table \ref{table}). 

In our work we consider the $f_1(1285)$-meson as a pure non-strange state ($\phi=0$). Presently there are some indications that such mixing in the axial-vector nonet is really small and $f_1(1285)$ is mostly made of $u$ and $d$ quarks. For instance, LHCb Collaboration \cite{Aaij14} gives $\phi =\pm (24.0^{+3.1+0.6}_{-2.6-0.8})^\circ$, assuming that $f_1(1285)$ is mixed with the $f_1(1420)$ state. This agrees with an earlier determination of $\phi =(-15^{+5}_{-10})^\circ$ in \cite{Gidal87}. 
        
\begin{table}
\caption{NJL-model predictions for anomalous radiative decays of $f_1(1285)$ and $a_1(1260)$ axial-vector mesons, $\Gamma_{\mbox{\tiny NJL}}$. We give also some known empirical data, $\Gamma_{\mbox{\tiny exp}}$, and predictions of the covariant oscillator quark model \cite{Ishida89}. All decay widths are given in keV.  
\\}
\label{table}
\begin{tabular}{ccccc}
\hline
  Mode 
& $f_1\!\to\!\rho^0\gamma$
& $f_1\!\to\!\omega\gamma$
& $a_1\!\to\!\omega\gamma$
& $a_1\!\to\!\rho\gamma$ \\
\hline
 $\Gamma_{\mbox{\tiny NJL}}$ 
& 311 
& 34.3 
& 238 
& 26.8 \\
  $\Gamma_{\mbox{\tiny exp}}$ 
& $453\pm 177$ \cite{Dickson16}
&  
&   
&  \\

& $675\pm 313$ \cite{Amelin95}
&  
&   
&  \\

& $1326\pm 313$ \cite{PDG}
&  
&   
&  \\

  $\Gamma_{\mbox{\tiny mod}}$ \cite{Ishida89}
& 509-565  
& 48-57   
& 537  
& 62    \\

\hline
\end{tabular}
\end{table}

Calculating the triangle diagrams we considered the lowest order terms in an external momenta expansion (minimal couplings). It means that we are only concerned with the part of the effective action having the smallest possible number of derivatives which is responsible for the intrinsic parity violating processes. This approximation corresponds to the standard counting of the spin-1 mesons in Resonance Chiral Theory. In that approach, they contribute at ${\cal O}(p^6)$ order of chiral counting in the effective meson Lagrangian (i.e. with terms kept up to four derivatives). For further arguments supporting this approximation, we refer to the original papers \cite{Ecker89,Gasser89,Donoghue89}. Nonetheless, due to the importance of the question, we give some additional arguments in the following two sections.

It is worth to be mentioned here the quark model predictions by Lakhina and Swanson (see ref. [52] in \cite{Dickson16}). They have found that a nonrelativistic Coulomb-plus-linear quark potential model predicts $\Gamma_{f_1\to\gamma\rho^0}=1200\,\mbox{keV}$, while a relativized version of the model gives much less value $\Gamma_{f_1\to\gamma\rho^0}=480\,\mbox{keV}$. One sees that the NJL model result based on the relativistic quantum field theory calculations is in agreement with the relativized version of the Lakhina and Swanson model. Both models nicely reproduce the CLAS Collaboration result. On the other hand, the PDG-based estimate favors the nonrelativistic result.        

\section{Restrictions from QCD}
\label{pertnonpert}
The anomalous quark triangle diagrams considered here are the subject of special attention in the literature. The pioneering studies have been done by Rosenberg \cite{Rosenberg63} and Adler\cite{Adler69b}. Rosenberg got an explicit expression for the fermion triangle graph:
\begin{eqnarray}
\label{Tallq}
T_{\alpha\beta\gamma} (p,q) &=&\frac{1}{4\pi^2}\left\{e_{\alpha\beta\gamma\sigma}\left[q^\sigma(qp A_3+p^2 A_4) \right.\right. \nonumber \\
&-&\left. p^\sigma (qp A_3+q^2 A_4)\right] \nonumber \\
&-&e_{\alpha\gamma\rho\sigma}q^\rho p^\sigma (q_\beta A_3+p_\beta A_4) \nonumber \\
&+&\left. e_{\alpha\beta\rho\sigma}q^\rho p^\sigma (q_\gamma A_4+p_\gamma A_3)\right\},
\end{eqnarray} 
where we follow his notations
\begin{eqnarray}
&& A_3(p,q)=-16\pi^2 I_{11}(p,q), \\
&& A_4(p,q)=16\pi^2\left[I_{20}(p,q)-I_{10}(p,q)\right], \\
\label{st}
&& I_{st}(p,q)=  \\
&&\int_0^1\!\!\!dx\!\!\int_0^{1-x}\!\!\!\!\!\!\!\!\!dy\frac{x^sy^t}{x(1-x)p^2+y(1-y)q^2+2xy (qp)-m^2}. \nonumber
\end{eqnarray}
Here $m$ is a mass of the fermion in the triangle. The effective Lagrangian corresponding to (\ref{Tallq}) is, strictly speaking, local only if (\ref{st}) can be well approximated by $I_{st}(0,0)$. In this case, we obtain from above 
\begin{equation}
\label{As}
A_3(0,0)=\frac{2\pi^2}{3m^2}, \quad A_4(0,0)=\frac{4\pi^2}{3m^2}. 
\end{equation} 
This is exactly the approximation that has been used in our estimates of $T_{\alpha\beta\gamma} (p,q)$ (in our case $m$ is the constituent quark mass). Locality of the model requires us to work with form-factors $A_3 (p,q)$ and $A_4 (p,q)$ considered at $p^2=0$ and $l^2=0$, i.e. off mass-shell of variables $p^2$ and $l^2$.

It is, however, not well-understood what is the accuracy of such a step. Some interesting insight can be obtained by considering the amplitude in the limit of small external photon momentum $q$ \cite{Kuraev92}. In this limit the tensor $T_{\alpha\beta\gamma} (p,q)$ is linear in $q$ (one neglects quadratic and higher powers of $q$), and we obtain from (\ref{Tallq}) 
\begin{eqnarray}
\label{Tlinq}
T_{\alpha\beta\gamma} (p,q) &\!\!=\!\!&\frac{1}{4\pi^2}\left[A_4(p,0)(e_{\alpha\beta\gamma\sigma} p^2 + e_{\alpha\gamma\rho\sigma} p^\rho p_\beta)q^\sigma \right. \nonumber \\
&\!\!+\!\!&\left. A_3(p,0) (e_{\alpha\beta\rho\sigma}q^\rho p_\gamma - e_{\alpha\beta\gamma\sigma} qp) p^\sigma\right].
\end{eqnarray} 
Due to a special kinematics, which corresponds now to the decay of the axial-vector state in flight, the expressions for the form-factors $A_3(p,0)$ and $A_4(p,0)$ of the fermion triangle graph are considerably simplified
\begin{eqnarray}
\label{A34}
&& A_3(p,0)=-8\pi^2\int_0^1\!\!\!dx\frac{x(1-x)^2}{x(1-x)p^2-m^2}, \nonumber \\
&& A_4(p,0)= 2A_3(p,0),   
\end{eqnarray}  
but still follow the pattern $A_4(0,0)=2A_3(0,0)$. In general these two form-factors are independent.  

The equation (\ref{Tlinq}) can be cast into the standard form with the aid of the Schouten's identity \cite{Schouten38}
\begin{eqnarray}
&&(fa)|bcde|+(fb)|cdea|+(fc)|deab|+(fd)|eabc|\nonumber \\
&&+(fe)|abcd|=0,
\end{eqnarray}
where $|abcd|\equiv a_\mu b_\nu c_\alpha d_\beta e^{\mu\nu\alpha\beta}$. This identity allows us to write 
\begin{eqnarray}
\label{Tlinqmod}
&&T_{\alpha\beta\gamma} (p,q) =\frac{p^\sigma q^\rho}{4\pi^2}\left[w_L(p^2)\, p_\alpha e_{\beta\gamma\sigma\rho} \right.  \nonumber \\
&&-\left. w_T(p^2) \left( -p_\sigma e_{\alpha\beta\gamma\rho} +p_\alpha e_{\beta\gamma\sigma\rho} +p_\beta e_{\gamma\alpha\sigma\rho}\right)\right],
\end{eqnarray}
where the invariant functions $w_L(p^2)\equiv A_4(p,0)$, and $w_T(p^2)\equiv A_4(p,0)-A_3(p,0)$ are the longitudinal and transversal parts of the quark triangle with respect to the axial-vector momentum $l^\alpha$. Both structures are transversal with respect to the photon momentum $q^\gamma$, $q^\gamma T_{\alpha\beta\gamma =0}$. 

Let us consider now the problem from a different angle, namely, by using the one-quark-loop QCD result for the triangle graph. In this case, the longitudinal and transversal form-factors are still given by the equations (\ref{A34}), where one should only replace the constituent quark mass $m$ by the current quark mass $\hat m$. The result is (for $p^2\geq 4\hat m^2$)
\begin{eqnarray}
\label{QCD}
&& w_L(p^2)=2w_T(p^2)= -\frac{8\pi^2}{p^2} \nonumber \\
&& \left[1+\frac{2\hat m^2}{p^2\sqrt{1-\frac{4\hat m^2}{p^2}}}\left(\ln\frac{1+\sqrt{1-\frac{4\hat m^2}{p^2}}}{1-\sqrt{1-\frac{4\hat m^2}{p^2}}}-i\pi\right)\right] 
\end{eqnarray}   
In particular, in the chiral limit, $\hat m\to 0$, one gets 
\begin{equation}
\label{wL}
w_L(p^2)=2w_T(p^2)= -\frac{8\pi^2}{p^2+i\epsilon}, \quad (\hat m=0).
\end{equation} 

The longitudinal part represents the axial anomaly associated with the divergence of the axial-vector current $J_\alpha^5=\bar q\gamma_\alpha\gamma_5Aq$ (where $A=\tau_0/2$ for the $f_1$ meson case, or $A=\tau_3/2$ for the $a_1$ meson case) constructed from the light quark fields. Indeed,
\begin{eqnarray}
\label{axan}
&&\partial^\alpha J_\alpha^5[\hat m=0]\sim l^\alpha T_{\alpha\beta\gamma}\epsilon^\beta (p)\epsilon^\gamma (q)=
\nonumber \\
&&\frac{p^2}{4\pi^2}w_L(p^2) e_{\beta\gamma\sigma\rho}\epsilon^\beta (p)\epsilon^\gamma (q)p^\sigma q^\rho =\rho^{\sigma\beta}\tilde F_{\sigma\beta}, 
\end{eqnarray}  
where $\rho^{\sigma\beta}=p^\sigma\epsilon^\beta-p^\beta\epsilon^\sigma$, and $\tilde F_{\sigma\beta}=\frac{1}{2}e_{\sigma\beta\rho\gamma} F^{\rho\gamma}$. Note, that the imaginary part of $\partial^\alpha J_\alpha^5[\hat m=0]$ is equal to zero since $p^2\delta (p^2)=0$ \cite{Dolgov71}.

The Adler-Bardeen theorem \cite{Adler69} implies that the one-loop result (\ref{wL}) for $w_L(p^2)$ stays intact when the interaction with gluons is switched on. It is not corrected at the nonperturbative level too. Moreover, for the special kinematics considered (the photon carries a soft momentum) the transversal part $w_T(p^2)$ is unambiguously fixed by the longitudinal one $w_L(p^2)=2w_T(p^2)$ in the chiral limit of perturbation theory (up to the nonperturbative corrections to $w_T(p^2)$) \cite{Vainshtein03}.

The model result (\ref{T}) taken in special kinematics can be compared with these general QCD statements. Indeed, one can see that (a) it follows the QCD pattern $w_L=2w_T$; (b) it does not contain the imaginary part, which otherwise would contribute in the chiral limit $\hat m\to 0$ to the divergence of the axial-vector current; (c) it correctly reflects the underlying anomaly. The latter needs some explanation. On the mass-shell of the $\rho$-meson, we have         
\begin{equation}
\label{mda}
l^\alpha T_{\alpha\beta\gamma}\epsilon^\beta (p)\epsilon^\gamma (q)=-\frac{m_\rho^2}{6m^2}\rho^{\sigma\beta}\tilde F_{\sigma\beta}.
\end{equation} 
Numerically the factor $m_\rho^2/6m^2=1.28$ agrees quite well with factor 1 in (\ref{axan}). (The difference in the overall sign is not essential here because the sign in (\ref{axan}) can be changed by the appropriate definition of the QCD AVV-currents correlator.) But it is valid to expect from (\ref{axan}) and (\ref{mda}) of the full agreement, because we are dealing with the anomaly. And this expectation is actually correct. To show this let us use the mass formulae of the model
\begin{equation}
\frac{m_\rho^2}{6m^2}=\frac{1}{Z-1},\qquad Z=\frac{m_{a_1}^2}{m_\rho^2}.
\end{equation}
It follows then, that the value $Z=2$ would be in a total agreement with QCD requirement (\ref{axan}). Exactly this value of $Z$ was obtained long ago on the basis of spectral-function sum rules \cite{Weinberg67}, which are valid in QCD for $\hat m=0$, and of the KSFR relation for the $\rho$ coupling to the isospin current \cite{Suzuki66,Riazuddin66}. Now, we come to the same conclusion on the basis of the anomaly consideration.  

This reasoning, however, requires a caveat. From the QCD calculations, it follows that the axial anomaly has a pole at $p^2=0$. This pole implies the presence of zero-mass bound states in the physical spectrum \cite{Frishman81,Achasov93}. In the case of $a_1$ radiative decays, the $1/p^2$ pole in $w_L(p^2)$ can be associated with pions. However, for the $f_1$ case, there is no such a light pseudoscalar state (the $U(1)$ problem). For further progress with the singlet case, the $U(1)$ gluon anomaly must be included. Some interesting attempts may be found in \cite{Efremov90,Teryaev90,Teryaev95}, where the authors argue that the photon anomaly must be canceled by the gluon one. The question has been also addressed in the instanton liquid model of QCD vacuum \cite{Dorohov05}, where it has been shown that the singlet longitudinal amplitude $w_L^{(s)}(p^2)$ is renormalized at low momenta by the presence of the $U(1)$ gluon anomaly. As a result, the product $p^2 w_L^{(s)}(p^2)$ vanishes at $p^2=0$, taking the value $p^2 w_L^{(s)}(p^2)\simeq 0.8$ on the $\rho$-meson mass-shell. On the contrary, the normalized nonsinglet amplitude $w_L^{(ns)}(p^2)$ follows the pattern $p^2 w_L^{(ns)}(p^2)=1$ (this value exactly corresponds to the factor 1 in eq.(\ref{axan})). This shows that inclusion of the gluon anomaly will slightly change our estimates for the singlet $f_1$ decays. Nonetheless, the deviation from the equality $w_L^{(ns)}(p^2)=w_L^{(s)}(p^2)$ used in our calculations is rather small for the physical region of $p^2=m_\rho^2\simeq m_\omega^2$ considered here. 

\section{The $f_1\to\gamma\gamma$ results}
\label {f1ggampl}
It is straightforward now to obtain in the approximation considered above the  $f_1(1285)\to\gamma^*\gamma^*$ amplitude. The result is
\begin{equation}
{\cal M}_{f_1\gamma*\gamma^*}=T^{\mu\nu\alpha}_{f_1\gamma^*\gamma^*}(q_1,q_2)\epsilon^*_\mu (q_1) \epsilon^*_\nu (q_2)\epsilon_\alpha (l),
\end{equation} 
where $q_1$, $q_2$ are four-momenta of photons, $l$ is a four-momentum of the $f_1(1285)$ meson, $\epsilon_\mu (q_1)$, $\epsilon_\nu (q_2)$, $\epsilon_\alpha (l),$ are polarization vectors  and 
\begin{eqnarray}
&&T^{\mu\nu\alpha}_{f_1\gamma^*\gamma^*}(q_1,q_2)=-i\frac{5\alpha g_\rho}{36\pi m^2}\left\{ e^{\mu\nu\sigma\alpha}\left[q_{1\sigma}\left(q_1q_2+2q_1^2\right) \right.\right. \nonumber \\
&&-\left.q_{2\sigma}\left(q_1q_2+2q_2^2\right)\right]+e^{\rho\sigma\nu\alpha}q_{2\rho} q_{1\sigma} \left(q_2+2q_1\right)^\mu  \nonumber \\
&& +\left. e^{\rho\sigma\mu\alpha}q_{1\rho} q_{2\sigma} \left(q_1+2q_2\right)^\nu\right\}.
\end{eqnarray} 
In particular, in the rest frame of $f_1$, where $q_1=-q_2=k$, we obtain
\begin{eqnarray}
\label{ffgg}
&&T^{\mu\nu\alpha}_{f_1\gamma^*\gamma^*}(k,-k)=i\frac{5\alpha g_\rho}{18\pi m^2}e^{\mu\nu\alpha\sigma}k^2 k_\sigma \nonumber \\ 
&&\equiv8\pi i\alpha  e^{\mu\nu\alpha\sigma}k^2 k_\sigma F^{(0)}_{AV\gamma^*\gamma^*}(m_f^2, k^2,k^2).
\end{eqnarray}

Although the decay $f_1\to\gamma\gamma$ of a spin-one resonance is suppressed for real photons, according to the Landau-Yang theorem \cite{Landau48}, the value $F^{(0)}_{AV\gamma^*\gamma^*}(m_f^2, 0,0)$ of the matrix element (\ref{ffgg}) is measured with a good accuracy \cite{Aihara88,Achard02}.  For instance, an estimate \cite{Kochelev17} based on L3 Collaboration data gives
\begin{equation}
F^{(0)}_{AV\gamma^*\gamma^*}(m_f^2, 0,0)=(0.266\pm 0.043)\,\mbox{GeV}^{-2}.
\end{equation} 
A similar result one obtains from PDG \cite{PDG} values  
\begin{equation}
F^{(0)}_{AV\gamma^*\gamma^*}(m_f^2, 0,0)=(0.234\pm 0.034)\,\mbox{GeV}^{-2}.
\end{equation} 
Our estimate,
\begin{equation}
F^{(0)}_{AV\gamma^*\gamma^*}(m_f^2, 0,0)=\frac{5g_\rho}{(12\pi m)^2}=0.276\,\mbox{GeV}^{-2},
\end{equation} 
agrees well with these phenomenological values and gives an additional argument in favour of approximations made in the previous sections.

\section{Conclusions}
\label{concl}
In this work, we evaluate the radiative decay widths of the $f_1(1285)$ and $a_1(1260)$ axial-vector mesons with the assumption that these states are mostly made of $u$ and $d$ quarks. The anomalous triangle diagrams are considered. We show that surface term of these diagrams are fixed by the vector Ward identity and Bose statistics. These leads to the total cancellation of the linear in external momenta contributions in the low-energy expansion of the amplitudes. As a result, the leading order contribution (in the external momenta expansion) is determined by the cubic in momenta terms. This finite contribution in the framework of NJL model is totally fixed by the coupling of the $\rho\to\pi\pi$ decay, $g_\rho$, the finite structure constant $\alpha =1/137$, and masses of light quarks $m_u=m_d=280\ \mbox{MeV}$ and mesons. Note, that the constituent quark mass in the NJL model can be related with the phenomenological parameters only. The relation is given by the formula
\begin{equation}
6 \left(\frac{m}{g_\rho f_\pi}\right)^2\left(1-\frac{6m^2}{m_{a_1}^2}\right)=1.  
\end{equation}

The effective Lagrangian describing the radiative decays of these mesons are obtained. 

The decay width found, $\Gamma_{f_1\to\rho\gamma}=311\ \mbox{keV}$, is compatible with the recently measured value $\Gamma_{f_1\to\rho^0\gamma}=453\pm 177\ \mbox{keV}$ within errors, and four times less the value, quated by the Particle Data Group. Further study of the $f_1(1285)$ decay modes seems called for. The main point learned here is that the radiative decays of $f_1$ and $a_1$ mesons are very restricted by the anomalous character of these interactions.  

The method used here can be easily extended to study the radiative decays of the first radial excitations of the main axial-vector nonet. In particular, the model \cite{Volkov97,Volkov97a} can be used for that. Another interesting application of our result could be the study of $\gamma^*\to f_1(1285)\gamma$ mode in the $e^+e^-$ beams. The Lagrangians suggested can be also checked of mass-shell in the study of $\tau$-lepton decays, e.g. $\tau\to\nu_\tau\omega\rho$, or $\tau\to\nu_\tau\omega\pi^-\pi^0$.  \\              

\noindent {\it Acknowledgments:} We are indebted to O. V. Teryaev, M. A. Ivanov and N. I. Kochelev for important remarks which improved the paper.


\end{document}